\documentclass{elsart}
\usepackage{natbib}
\usepackage{epsfig}
\begin{document}
\runauthor{Cicero, Caesar and Vergil}
\begin{frontmatter}
\title{The Fermi motion contribution to $J/\psi$ production at
the hadron colliders }
\author{M.A. Gomshi Nobary$^{a,b}$ and B. Nikoobakht$^{a}$}
\thanks[]{E-mail: mnobary@razi.ac.ir}
\address{$^a$ Department of Physics, Faculty of Science, Razi University,Kermanshah, Iran.}
\address{$^b$ The Center for Theoretical Physics and Mathematics, A.E.O.I.,
Roosbeh Building, P.O. Box 11365-8486 Tehran, Iran. }
\begin{abstract}
We investigate the relativistic Fermi motion effect in the case of
$J/\psi$ production in various hadron colliders. A light-cone wave
function is adopted to represent the $J/\psi$ final state. The
change in the confinement parameter which sets a scale for the
size of the final state, allows one to see the effect in an
explicit manner. While the effect has considerable influence on
the fragmentation probabilities and the differential cross
sections, the total cross sections are essentially left unchanged.
such a feature is in agreement with the momentum sum rule which
the fragmentation functions should satisfy.

\end{abstract}
\end{frontmatter}

\section {Introduction}
Evaluation of the $J/\psi$ cross section at Tevatron energies has
been one of the interesting problems of QCD in theory [1,2,3] and
in experiment [4]. Predictions using the QCD calculations where
the production of $J/\psi$ is assumed to occur in color singlet
form, fails to agree with the experimental results. To bring about
the agreement the mechanism of color octet is introduced in which
the bound state is produced originally in the color octet state at
the production point [5]. Then by emitting a soft gluon the
colored object is transformed into a color singlet state [6].

Generally in the formation of quarkonia bound state it is assumed
that the constituents are not relativistic and therefore they are
let to fly together within the bound state ignoring their
respective motion. Here we propose that the constituents within
such states specially the charm quark is not that heavy to let one
to ignore its relativistic effects particularly in relation to the
energies at which such a particle is produced. Therefore the
effect should be accounted for in the production process in a more
explicit and accurate form. The results of such a study may shed
light on the problem of $J/\psi$ production cross section.

In this work, we introduce the Fermi motion into the $J/\psi$
production in direct fragmentation process using a light-cone wave
function. Since our aim is to show the size of the effect, we have
not included other contributions. We demonstrate the enhancement
of the fragmentation function due to this effect. With such a
significant enhancement, we evaluate the differential cross
section times the branching ratio and the total integrated cross
section for the process $\bar p p\rightarrow c\rightarrow
J/\psi\rightarrow \mu^+\mu^-$ at the Tevatron Run I energies and
compare them with the CDF data. We also present similar results
for the case of the RHIC, the Tevatron Run II and the CERN LHC
$pp$ collisions.

\section {The light-cone wave function and the bound state formation via fragmentation}

There have been different approaches to introduce the Fermi motion
into the production and decay processes of various meson states
and quarkonia using different forms of wave functions [7]. The
light-cone wave functions have been employed in the case of
charmonia to study their production in B-decay [8] and in
photoproduction [9]. Here motivated by harmonic oscillator model,
we have picked up a wave function in the light-cone quantization
to represent the $J/\psi$ bound state. It has the following form
[7]
\begin{eqnarray}
\psi_{c\rightarrow J/\psi}(x_1,x_2,q_T)=A_{c\rightarrow
J/\psi}\exp
\Biggl[-{1\over{8\beta^2}}{{m^2+q_T^2}\over{x_1x_2}}\Biggl],
\end{eqnarray}
where $A_{c\rightarrow J/\psi}$ is the normalization coefficient,
$m$ is the quark (anti-quark) mass and $q_T$ is the transverse
momentum of the constituents. The $x$'s are the energy momentum
ratios and finally the parameter $\beta$ is known as the
confinement parameter which controls the width of the wave packet
representing the bound state. The normalization condition is
\begin{eqnarray}
\sum_{n,\lambda_i} \int[dx][d^2{\bf q}_T]|\psi_n(x_i,{\bf
q}_{T_i},\lambda_i)|^2=1,
\end{eqnarray}
where
\begin{eqnarray}
[dx]\equiv \prod_{i=1}^n dx_i\delta\Bigl[1-\sum_{i=1}^n x_i\Bigr],
\end{eqnarray}
and
\begin{eqnarray}
[d^2 {\bf q}_T]\equiv \prod_{i=1}^n d^2{\bf q}_{T_i} 16\pi^3
\delta^2\Bigr[\sum_{i=1}^n {\bf q}_{T_i}\Bigl].
\end{eqnarray}
The sum is over all Fock states and helicities.

With the choice of the wave function (1), we fix all degrees of
freedom (transverse and longitudinal) of the constituents within
the $J/\psi$ bound state by matching them with those in the matrix
elements relevant to the fragmentation function and vary the
parameter $\beta$ to change the size of the wave packet
representing the bound state. Benefiting such a method, in the
leading order perturbative regime, the fragmentation functions for
$J/\psi$ production without and with the Fermi motion are obtained
as follows [10]:

\noindent {\bf (a) Fermi motion off}

\noindent In this case the transverse momenta of the constituents
are set equal to zero and that the longitudinal components are
chosen to be equal. The confinement parameter is $\beta=0$ in this
case. The fragmentation function is obtained as

\begin{eqnarray}
D_{c\rightarrow J/\psi}(z,\mu_\circ,\beta=0)& =&{{\alpha_s^2
C_F^2\langle {k_T}^2\rangle^{1/2}}\over 16 m { F(z)}}\biggl\{
z(1-z)^2\Bigl[\xi^2z^4+2\xi z^2(4-4z+5z^2)\nonumber\\
&&+(16-32z+24z^2-8z^3+9z^4)\Bigr]\biggr\},
\end{eqnarray}
where $\alpha_s$ is the strong interaction coupling constant and
$C_F$ is the color factor. The quantity $ \langle {k_T}^2\rangle$
is the average transverse momentum squared of the initial state
heavy quark, the parameter $\xi$ is defined as $ \xi=\langle
{k_T}^2\rangle/m^2 $ and finally $F(x)$ is given by

\begin{eqnarray}
{ F(z}) =\Bigl[\xi^2z^4-(z-2)^2(3z-4)+\xi z^2(8-7z+z^2)\Bigr]^2.
\end{eqnarray}

\noindent {\bf (b) Fermi motion on}

Here the different momentum components of the constituents, i.e.
the quantities $q=|{\bf q}_T|$ and $x=x_1=1-x_2$ in the matrix
elements and the wave function (1) are integrated over. The only
remaining parameter is the confinement parameter $\beta$. The
fragmentation function in this case is obtained as

\begin{eqnarray}
D_{c\rightarrow J/\psi}(z,\mu_\circ,\beta)&
=&{{\pi^2\alpha_s^2C_F^2\langle {k_T}^2\rangle^{1/2}}\over {2 m
}}\int \frac{dq dx
|\psi_M|^2x^2(1-z)^2 zq}{{ G(z})}\nonumber\\
&&\times\Bigl\{1-4(1-x)z+2(4-10x+7x^2)z^2\nonumber\\
&&+4(-1+x^3-5x^2+4x)z^3+(1-4x+8x^2-4x^3+x^4)z^4\nonumber\\
&&+\eta\xi z^2\bigl[1-2x+z^2+x^2(2-2z+z^2)\bigr]+ \eta\bigl[
2+(-6+4x)z\nonumber\\
&&+(9-8x+2x^2)z^2-2(2-x+x^2)z^3+(1+x^2)z^4\bigr]\nonumber\\
&&+\xi z^2\bigl[1+2x^3(2-3z)z+z^2+2x^4z^2+2x(-1+z-2z^2)\nonumber\\
&& +x^2(2-8z+9z^2)\bigr]+\eta^2(1-z)^2+\xi^2(1-x)^2x^2z^4\Bigr\}.
\end{eqnarray}
The function ${ G(z)}$ reads
\begin{eqnarray}
{ G(z)}&=&\Bigl\{\bigl[\eta (1-z)^2+\xi
x^2z^2+(1-(1-x)z)^2\bigr]\nonumber\\ &&\times\bigl[\eta(-1+z)+\xi
(-1+x)xz^2-1+(1-x+x^2)z\bigr]\Bigr\}^2.
\end{eqnarray}
Here we have defined $\eta=q^2/m^2$. In our study we have set the
charm quark mass equal to 1.25 GeV.

The fragmentation functions (5) and (7) provide a comparison of
the cases of the Fermi motion off and on. The two functions
coincide at sufficiently low $\beta$. As $\beta$ increases, (7)
raises considerably at the peak region which leads to the
respective increase in fragmentation probability. To choose the
$\beta$ value, we note that different values within the range of
$\beta=$0.250 - 0.750 GeV have been employed for different meson
states [7]. Even the case of $\beta=m_q$ is adopted by Tao Huang
et al. in [7]. On the other hand in [10], we have argued that the
size of the wave packet representing the bound state is important
here. Therefore we have raised the value of $\beta$ up to 0.6 GeV
which we believe is quite safe. The behavior of the fragmentation
function (7) is shown in Fig. 1 for $\beta=0,\; 0.2, \;0.4 \;{\rm
and} \;0.6$ GeV. We take the maximum value of $\beta$=0.6 GeV and
use it for our further considerations.

\begin{figure}
\begin{center}
\includegraphics[width=10cm]{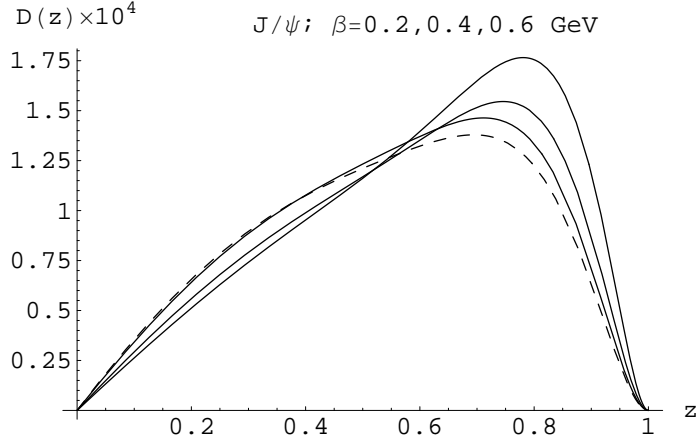}
\caption{Fragmentation function for $J/\psi$ production. While the
dashed curve represents the function when the Fermi motion is off,
the solid ones show the behavior of this function when the Fermi
motion is on. The $\beta$ values are indicated. The function picks
up as $\beta$ increases. This gives raise to increase in the
fragmentation probability. }
\end{center}
\end{figure}

\section {Inclusive production cross section }

We have employed the idea of factorization to evaluate the
$J/\psi$ production cross section at hadron colliders. For $\bar
{p} p$ collisions we may write

\vfil\eject

\begin{eqnarray}
\frac {d\sigma}{dp_T}(\bar{p} p \rightarrow c \rightarrow J/\psi
(p_T) X)&=&\sum_{i,j}\int
dx_1 dx_2 dz f_{i/\bar{p}}(x_1,\mu)f_{j/ p}(x_2,\mu)\nonumber\\
&&\times\Bigl[ \hat\sigma(ij\rightarrow c(p_T/z)X,\mu) D_{
c\rightarrow J/\psi}(z,\mu,\beta)\Bigr].
\end{eqnarray}

Where $f_{i,j}$ are parton distribution functions with momentum
fractions of $x_1$ and $x_2$ (different from $x_1$ and $x_2$ which
appear in (1)), $\hat\sigma$ is the charm quark production cross
section and $D(z,\mu,\beta)$ represents the fragmentation of the
produced heavy quark into $\bar c c$ state with confinement
parameter $\beta$ at the scale $\mu$. We have used the
parameterization due to Martin-Roberts-Stiriling (MRS) [11] for
parton distribution functions and have included the heavy quark
production cross section up to the order of $\alpha_s ^3$ [12].
The dependence on $\mu$ is estimated by choosing the transverse
mass of the heavy quark as our central choice of scale defined by
\begin{eqnarray}
\mu_R=\sqrt{ {p_T}^2{\rm (parton)}+{m_c}^2,   }
\end{eqnarray}
and vary it appropriate to the fragmentation scale of the particle
state to be considered. This choice of scale, which is of the
order of $p_T$ (parton), avoids the large logarithms in the
process of the form $\ln(m_Q/\mu)$ or $\ln(p_T/\mu)$. However, we
have to sum up the logarithms of order of $\mu_R/m_Q$ in the
fragmentation functions. But this can be implemented by evolving
them by the Altarelli-Parisi equation [13]. The following form of
this equation is used here

\begin{eqnarray}
\mu {\partial\over{\partial \mu}} D_{c\rightarrow J/\psi}
(z,\mu,\beta)= \int_z^1 {dy\over y} P_{Q \rightarrow Q} (z/y,\mu)
D_{c\rightarrow J/\psi}(y,\mu,\beta).
\end{eqnarray}
 Here $P_{Q \rightarrow Q}(x=z/y,\mu) $ is the Altarelli-Parisi
splitting function. The boundary condition on the evolution
equation (11) is the initial fragmentation function
$D_{c\rightarrow J/\psi}(z,\mu,\beta)$ at some scale
$\mu=\mu_\circ$. In principle this function may be calculated
perturbatively as a series in $\alpha_s$ at this scale.

Detection of final state requires kinematical cuts of the
transverse momentum, $p_T$, and the rapidity, $y$. We have imposed
the required $p_T^{\rm cut}$ and $y^{\rm cut}$ in our simulations
for different colliders as required and have used the following
definition of rapidity

\begin{eqnarray}
y=\frac{1}{2} \log\Biggl[\frac{E-p_L}{E+p_L}\Biggr].
\end{eqnarray}

\begin{figure}
\begin{center}
\includegraphics[width=14cm]{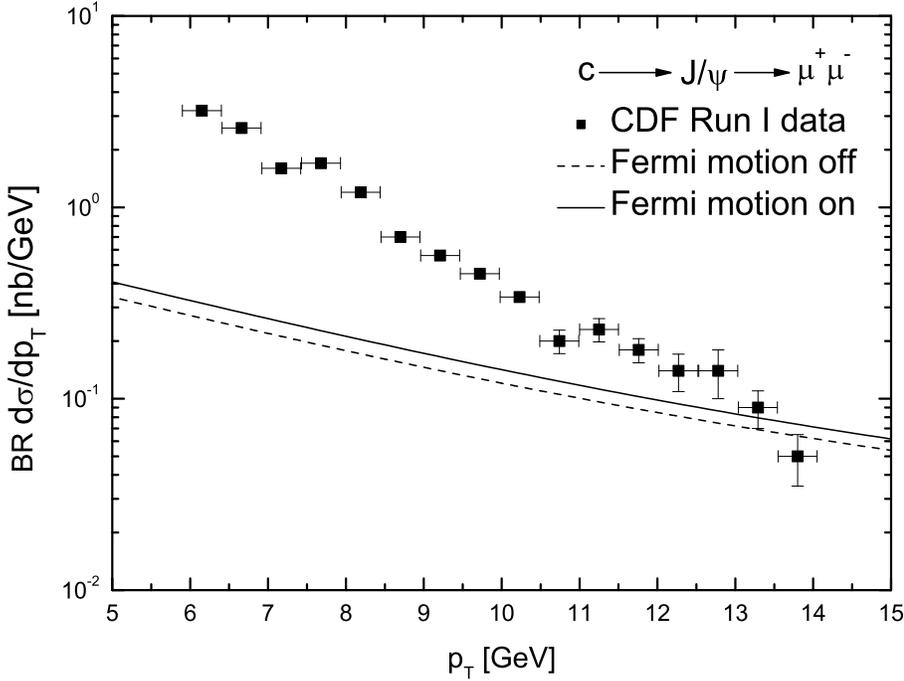}
\caption{The differential cross section for direct fragmentation
production of $J/\psi$ and its subsequent decay $J/\psi
\rightarrow \mu^+ \mu^-$ at the Tevatron Run I energies. While the
dashed curve is obtained using (5) or equally (7) with $\beta=0$,
the solid one is due to (7) with $\beta=0.6$ GeV. The result is
compared with the CDF Run I data. Other contributions are not
included. The scale is chosen to be $2\mu_R$. }
\end{center}
\end{figure}

\begin{figure}
\begin{center}
\includegraphics[width=14cm]{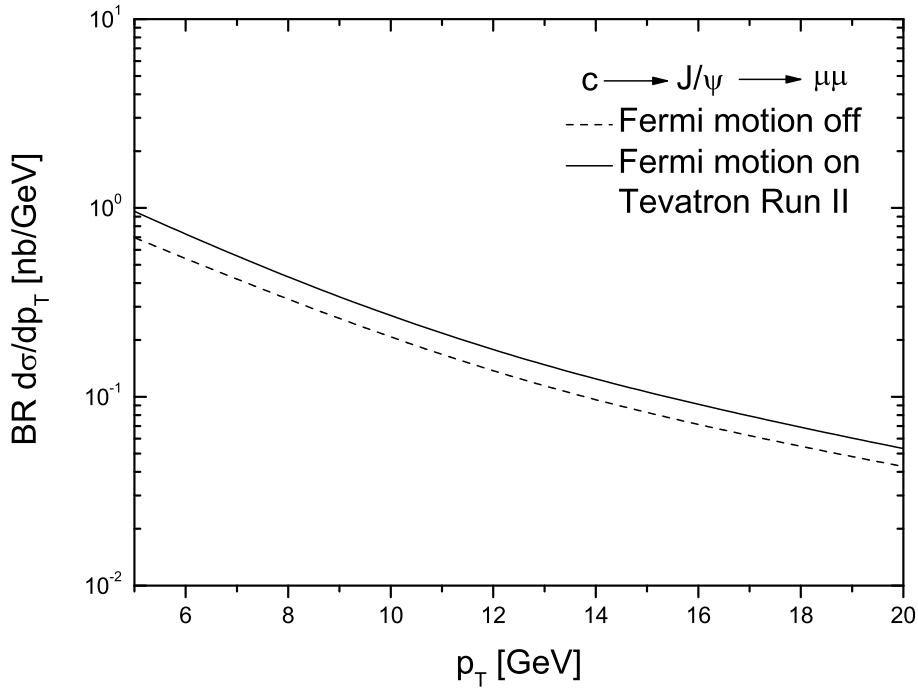}
\caption{The differential cross section for direct fragmentation
production of $J/\psi$ at Tevatron Run II. The two curves are
obtained using (7) with $\beta= 0$ and 0.6 GeV respectively. The
scale has been set to $2\mu_R$. }
\end{center}
\end{figure}

\begin{figure}
\begin{center}
\includegraphics[width=14cm]{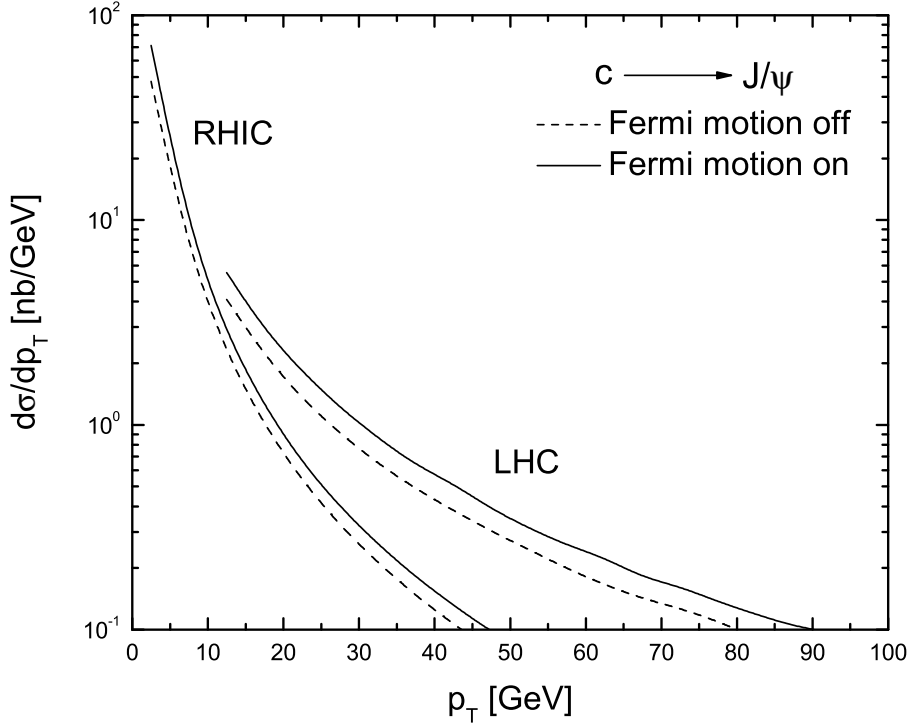}
\caption{The differential cross section for direct fragmentation
production of $J/\psi$ at the RHIC and the CERN LHC. The shift due
to the Fermi motion is significantly increased in the case of LHC
but shows to be less important at the RHIC. In both cases the
fragmentation function (7) is employed with $\beta= 0$ and 0.6 GeV
respectively.}
\end{center}
\end{figure}

\section {Results and discussion}

We have used a light-cone wave function to introduce the Fermi
motion in $J/\psi$ production in direct fragmentation channel and
obtained its fragmentation function in leading order perturbative
regime. In this function the confinement parameter switches the
effect of Fermi motion and within its physically acceptable
values, the function demonstrates the effect of Fermi motion in an
interesting manner. The motivation of introducing this effect is
to see its role in improving the QCD versus experimental results
for the $J/\psi$ cross section at the Tevatron energies.

In the case of the $J/\psi$ state such a fragmentation function,
with a reasonable choice of the confinement parameter, gives raise
to a significant increase in the fragmentation probability due to
the Fermi motion. Such a behavior is illustrated in Fig. 1.

To see the effect in the production rates, we have used the usual
procedure of factorization for $J/\psi$ production in hadron
colliders. We present the $p_T$ distribution of ${\rm BR}
(J/\psi\rightarrow\mu^+\mu^-)d\sigma/dp_T$ for the cases of the
Fermi motion off and on along with the CDF data at Run I in the
Fig. 2. The branching ratio BR($J/\psi\rightarrow
\mu^+\mu^-$)=0.0597 is taken from [14]. The poor agreement with
data is due to the fact that here we have only considered the
contribution of $\bar{p}p\rightarrow c\rightarrow J/\psi
\rightarrow \mu^+ \mu^-$. Similar behavior at Tevatron Run II
energies is shown in Fig. 3. We have also extended our study to
the cases of the RHIC and the CERN LHC $ p p$ colliders. Here we
provide the $p_T$ distributions of the differential cross sections
for $c\rightarrow J/\psi$ and compare the two cases of the Fermi
motion off and on in the Fig. 4. In all cases we have used
$\beta=0.6$ GeV for the confinement parameter in the fragmentation
functions. Naturally, the results for $\beta$ in the range of 0 -
0.6 GeV fall between the above results.

We have also calculated the total integrated cross sections for
each case. We found that the total cross sections for with and
without Fermi motion essentially remain unchanged within the
uncertainties of Mote Carlo simulations. The reason is first due
to the momentum sum rule which the fragmentation functions should
satisfy. In other words although the modification of fragmentation
functions by the Fermi motion redistributes the final states, the
integrated cross sections are left unchanged. Alternatively
although the Fermi motion increases the fragmentation probability
for the state, i.e., introduces a state with overall higher mass,
the cross section is lowered by just the same amount when we
introduce the effect in calculation of the total integrated cross
section.

First of all we note that the effect of Fermi motion is indeed
significant and that the implementation of such a study in all
production channels of $J/\psi$ may give raise to considerable
enhancement of the color singlet differential cross section.

It is evident from the Figures 2,3 and 4 that the effect increases
with increasing $\sqrt{s}$. The kinematical cuts play important
role apart from $\sqrt {s}$. The large cross section at the RHIC
compared with the LHC in the Fig. 4 is an example.

Finally we consider the uncertainties involved in our study. There
are two main sources of uncertainities.The first is about the
simulation of $J/\psi$ production at hadron colliders such as the
uncertainties along with the fragmentation functions and parton
distribution functions. These kind of uncertainties are well
discussed in the literature. The second source of uncertainty is
due to the choice of the confinement parameter in the
fragmentation function. Relying on our discussion in section 2,
our choice of $\beta=0.6$ GeV seems to be justified. Future
determination of this parameter will shed more light on this
situation. It is worth mentioning that our choice of charm quark
mass, i.e., 1.25 GeV, have put our results in their upper side and
that the change of the charm quark mass  in its acceptable range
does not have significant impact on the Fermi motion effect in
$J/\psi $ production.

\end{document}